\documentclass{aa}

\usepackage[utf8]{inputenc}
\usepackage{color}
\usepackage[usenames,dvipsnames,svgnames,table]{xcolor}
\usepackage{rotating}
\usepackage{txfonts}
\usepackage[]{hyperref}
\hypersetup{colorlinks=true,citecolor=blue}
\usepackage{multirow} 
\definecolor{ora}{RGB}{238,118,0}

\begin{document} 

   \title{Revised mass-radius relationships for water-rich rocky planets more irradiated than 
the runaway greenhouse limit.}

   \author{Martin Turbet \inst{1}, Emeline Bolmont\inst{1}, David Ehrenreich\inst{1}, 
Pierre Gratier\inst{2}, Jérémy Leconte\inst{2}, Franck Selsis\inst{2}, Nathan Hara\inst{1}, Christophe Lovis\inst{1}}

 \institute{Observatoire astronomique de l’Université de Genève, 51 chemin des Maillettes, 1290 Sauverny, Switzerland
 \email{martin.turbet@unige.ch} \and 
 Laboratoire d'astrophysique de Bordeaux, Univ. Bordeaux, CNRS, B18N, all\'ee Geoffroy Saint-Hilaire, 33615 Pessac, France}

\date{Received; accepted}

 \abstract{Mass-radius relationships for water-rich rocky planets are usually calculated 
assuming most water is present in condensed (either liquid or solid) form. 
Planet density estimates are then compared to these mass-radius relationships, 
even when these planets are more irradiated than the runaway greenhouse irradiation limit 
(around 1.1~times the insolation at Earth for planets orbiting a Sun-like star), 
for which water has been shown to be unstable in condensed form and would instead form a thick H$_2$O-dominated atmosphere. 
Here we use a 1-D radiative-convective inverse version of the LMD generic numerical climate model to 
derive new theoretical mass-radius relationships appropriate for water-rich rocky planets that are more irradiated than the 
runaway greenhouse irradiation limit, meaning planets endowed with a steam, water-dominated atmosphere.
As a result of the runaway greenhouse radius inflation effect introduced in previous work, 
these new mass-radius relationships significantly differ from those traditionally used in the literature. 
For a given water-to-rock mass ratio, these new mass-radius relationships lead to planet bulk densities much lower 
than calculated when water is assumed to be in condensed form. 
In other words, using traditional mass-radius relationships for planets that are more irradiated than the 
runaway greenhouse irradiation limit tends to dramatically overestimate -possibly by several orders of magnitude- 
their bulk water content. In particular, this result applies to TRAPPIST-1 b, c, and d, 
which can accommodate a water mass fraction of at most 2, 0.3 and 0.08$\%$, respectively, 
assuming planetary core with a terrestrial composition. 
In addition, we show that significant changes of mass-radius relationships (between planets 
less and more irradiated than the runaway greenhouse limit) can be used to remove bulk composition 
degeneracies in multiplanetary systems such as TRAPPIST-1.
Broadly speaking, 
our results demonstrate that non-H$_2$/He-dominated atmospheres can have a first-order effect on 
the mass-radius relationships, even for rocky planets receiving moderate irradiation. 
Finally, we provide an empirical formula for the H$_2$O steam atmosphere thickness as a function of 
planet core gravity and radius, water content, and irradiation. This formula can easily be used to construct mass-radius relationships for any 
water-rich, rocky planet (i.e., with any kind of interior composition ranging from pure iron to pure silicate) 
more irradiated than the runaway greenhouse irradiation threshold.}

\titlerunning{Revised mass-radius relationships for water-rich rocky planets more irradiated than the runaway greenhouse limit.}
\authorrunning{Martin Turbet et al.}
\maketitle

\section{Introduction}
\label{section_introduction}

With the discovery of the nearby TRAPPIST-1 system \citep{Gillon:2016,Gillon:2017,Luger:2017}, 
we now have seven rocky planets in temperate orbits for which both 
radii \citep{Gillon:2017,Luger:2017,Delrez:2018} and masses \citep{Grimm:2018} have been measured 
with unprecedented accuracy for planets of this nature.
\citet{Grimm:2018} and \citet{Dorn:2018} compared TRAPPIST-1 planets' 
bulk density\footnote{The densities of TRAPPIST-1 planets were measured with the transit timing variations (TTVs) technique. 
They are therefore absolute densities, and are thus not affected by inaccuracy on the stellar mass and radius measurements \citep{Grimm:2018}.} 
estimates with mass-radius relationships of rocky planets 
endowed with thick condensed water layers and inferred from the comparison that most of the seven planets are likely enriched in 
volatiles (e.g., water) up to several tens of percent of planetary mass.

We were motivated by these studies to recalculate mass-radius relationships for water-rich rocky planets in cases where all water 
is vaporized, forming a thick H$_2$O-dominated steam atmosphere. This situation has been shown to occur for planets receiving more 
irradiation from their host star than the theoretical runaway greenhouse irradiation limit \citep{Kasting:1993,Goldblatt:2012,Kopparapu:2013}. 
In the TRAPPIST-1 system, the three innermost planets (TRAPPIST-1 b, c, and d) are thought to receive more irradiation than the 
theoretical runaway greenhouse irradiation limit for ultra-cool stars \citep{Kopparapu:2013,Wolf:2017,Turbet:2018aa}, even when 
considering the possible negative feedback of substellar water clouds \citep{Yang:2013,Kopparapu:2016} 
expected on tidally locked planets.

Traditionally, mass-radius relationships \citep{Seager:2007,Sotin:2007,Grasset:2009,Mordasini:2012,Swift:2012,Zeng:2013,Zeng:2016} 
for water-rich rocky planets are calculated assuming water is either in solid or liquid form, depending on planet 
equilibrium temperatures. Some studies \citep{Dorn:2018, Zeng:2019} included the effect of a H$_2$O-rich atmosphere on the 
planetary radius estimate by assuming an isothermal steam atmosphere at the equilibrium planet temperature. 
\citet{Thomas:2016} explored the effect that a thick H$_2$O atmosphere may have on the mass-radius relationships of 
Earth to super-Earth-mass planets. To do this, they used a structural model forced at various surface temperatures and in various pressure-boundary conditions. 
Their model takes convection processes into account, but lacks a radiative transfer. 
As a result, the surface temperatures assumed in \citet{Thomas:2016} are significantly lower than those 
calculated self-consistently in the standard atmospheric numerical simulations taking into account 
the radiative exchanges both in short-wave and long-wave ranges \citep{Kopparapu:2013,Goldblatt:2013,Turbet:2019aa}. 
Radiative transfer is a necessary component to ensure that atmospheric states have reached top-of-atmosphere radiative balance, 
and thus describe physically realistic planets.

All the aforementioned approaches most likely underestimate the 
physical size of a H$_2$O-dominated steam atmosphere for planets receiving more irradiation than the runaway greenhouse limit 
\citep{Turbet:2019aa}. Using a 1-D numerical radiative-convective climate model, \citet{Turbet:2019aa} in fact recently showed that 
water-rich planets receiving more irradiation than the runaway greenhouse irradiation threshold should suffer from a strong 
atmospheric expansion compared to planets receiving less irradiation than this threshold. The effect, 
which they named the runaway greenhouse 
radius inflation effect, originates from the cumulative effect of four distinct causes: 
(i) a significant increase in the total atmospheric mass; 
(ii) a significant increase in the atmospheric temperatures; (iii) an increase in optical thickness at low atmospheric pressure;
and (iv) a decrease in the mean molecular mass.

To the best of our knowledge, the study of \citet{Valencia:2013} is the only work ever to 
have self-consistently considered the effect of a steam H$_2$O atmosphere on mass-radius relationships 
in the Earth to super-Earth mass regime planets. 
However, this work focused on highly irradiated planets only (around 20 times the insolation at Earth), 
with the aim of improving our understanding of the nature of the exoplanet GJ 1214b. 
Although they did not directly calculate mass-radius relationships, 
\citealt{Nettelmann:2011} (based on previous results from \citealt{Miller-Ricci:2010}) 
also carried out interior-atmosphere calculations self-consistently taking
into account the effect of a H$_2$O-dominated steam atmosphere to evaluate the 
possible nature of GJ~1214b. 
The results of \citet{Nettelmann:2011} and \citet{Valencia:2013} are qualitatively in agreement 
(and quantitatively in agreement in the case of GJ~1214b), 
that planets endowed with a steam H$_2$O atmosphere have a significantly larger radius than icy or liquid ocean planets, for a given water-to-rock ratio.
\citet{Thomas:2016} also recovered qualitatively similar 
results, that planets endowed with a steam H$_2$O atmosphere have a 
significantly larger radius than icy or liquid ocean planets, for a given water-to-rock ratio.

Here we make use of the 1-D inverse radiative-convective model previously introduced in \citet{Turbet:2019aa}, 
coupled to mass-radius relationships of rocky interiors from \citet{Zeng:2016}, 
to produce revised mass-radius relationships for rocky planets in temperate orbits endowed with thick H$_2$O 
steam envelopes, as predicted for water-rich planets receiving more irradiation than the runaway greenhouse limit \citep{Turbet:2019aa}. 
While \citet{Turbet:2019aa} focused on the theoretical and numerical ground of the runaway greenhouse 
radius inflation, as well as observational tests to detect it in the exoplanet population, 
here we derive and make available to the community mass-radius relationships aimed at 
better interpreting the nature of terrestrial-size planets, for which we are beginning to 
have increasingly accurate measurements of masses and radii.

In Section~\ref{section_methods}, we describe the method we used to 
calculate mass-radius relationships for planets endowed with steam, water-dominated atmospheres. 
These new mass-radius relationships are then presented and discussed in Section~\ref{section_results}. 
Lastly, we present the conclusions of this work and 
discuss future perspectives in Section~\ref{section_conclusions_discussions}.

\section{Methods}
\label{section_methods}

In this section, we describe first the method we used to calculate
mass-radius relationships for planets endowed with steam, 
water-dominated atmospheres. We then provide the empirical mass-radius relationship fitted to these calculations.

\subsection{Procedure to derive revised mass-radius relationships}

\begin{figure*}
    \centering
\includegraphics[width=\linewidth]{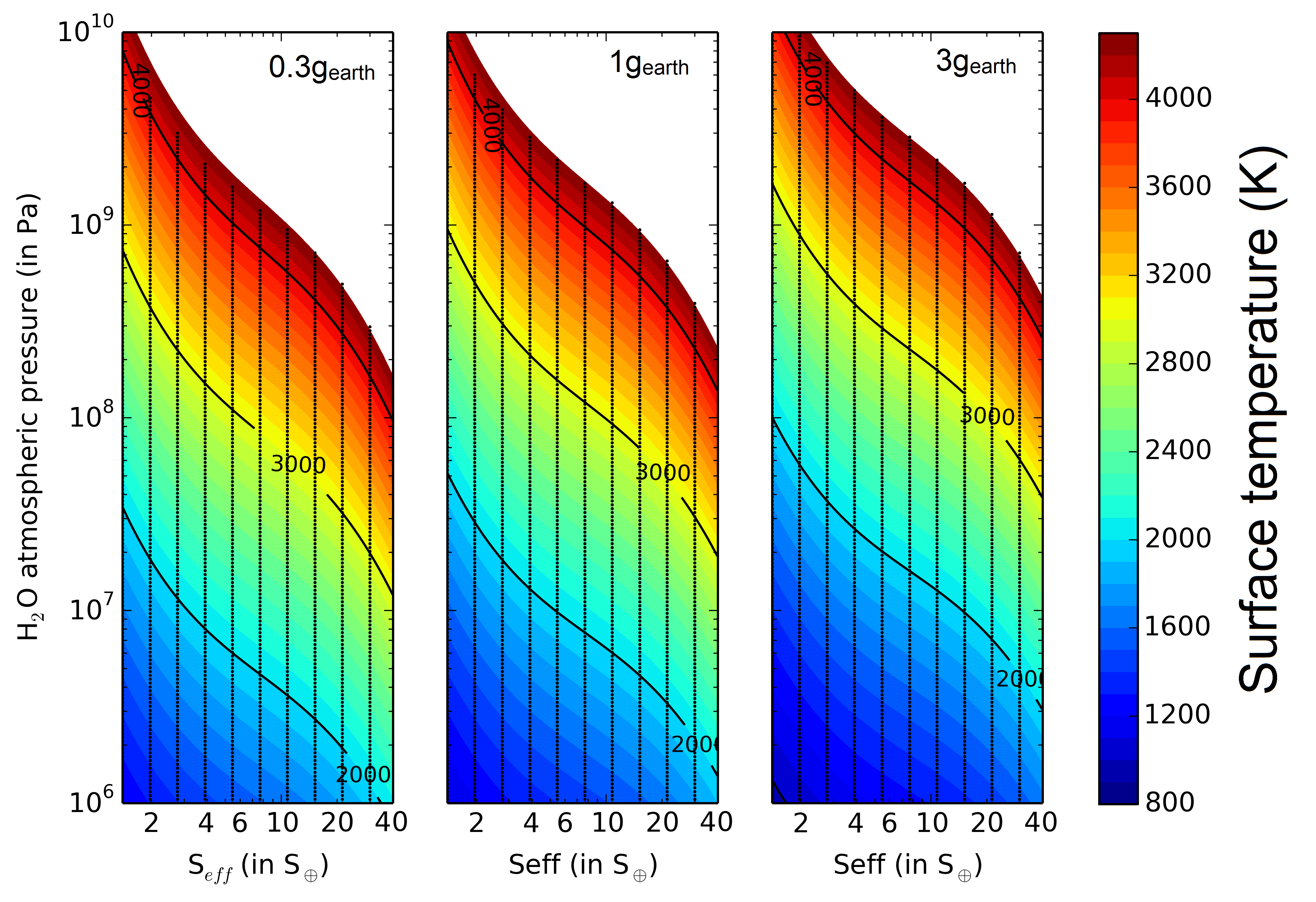}
\caption{Surface temperature as a function of the effective flux received on a planet (x-axis) 
and the surface pressure of its steam H$_2$O atmosphere (y-axis), for three different surface gravities (0.3, 1 and 3$\times$ the 
gravity on Earth). The surface temperature was estimated using Equation~\ref{empirical_equation_tsurf}. 
The small black dots indicate the parameter space for which atmospheric numerical simulations 
have been carried out, and on which the fit of the surface temperature is based.}
\label{tsurf_contourplot}
\end{figure*}

We calculated the mass-radius relationships for water-rich rocky planets that are more irradiated than the 
runaway greenhouse irradiation limit in four main steps: 
Firstly, we retrieved mass-radius relationships of dry, rocky planets. 
In this paper, we chose to use the mass-radius relationships of \citet{Zeng:2016}\footnote{User-friendly data is provided on the personal website 
of Li Zeng (\url{https://www.cfa.harvard.edu/~lzeng/planetmodels.html}).} for 
(i) pure silicate (MgSiO$_3$) planets; (ii) terrestrial core composition planets; and (iii) pure-iron (Fe) planets. However, any 
type of rocky interior composition (from pure iron to pure silicate) could be used.

Secondly, for each set of rocky interior mass M$_{\text{core}}$ and radius R$_{\text{core}}$, 
we calculated the transit thickness z$_{\text{atmosphere}}$ 
and the mass M$_{\text{atmosphere}}$ that a pure H$_2$O atmosphere would have 
for a wide range of possible water atmospheric pressures, 
using a 1-D inverse radiative-convective version of the LMD Generic model. 
The model was adapted in \citet{Turbet:2019aa} to simulate the vertical structure of steam atmospheres, taking into 
account the condensation of water vapor using a non-dilute moist lapse rate formulation as in \citet{Marcq:2017} and 
a radiative transfer using the water-dominated absorption coefficients of \citet{Leconte:2013nat}. 
For the calculation of the atmospheric profile, the change in gravity with altitude is also taken into account.
For more details on the model, we refer the reader to Appendix~A of \citet{Turbet:2019aa}.
This second step is discussed in more detail below. 

Thirdly, for each set of rocky interior mass and radius, and for each possible water atmospheric pressure, we calculated the 
resulting mass M$_{\text{planet}}$ and transit radius R$_{\text{planet}}$ by using 
M$_{\text{planet}}$~=~M$_{\text{core}}$+M$_{\text{atmosphere,}}$ and assuming that 
R$_{\text{planet}}$~=~R$_{\text{core}}$+z$_{\text{atmosphere}}$. In Appendix~\ref{appendix_trick_r-m}, we discuss in detail 
how the relationship R$_{\text{planet}}$~=~R$_{\text{core}}$+z$_{\text{atmosphere}}$ -- 
with R$_{\text{core}}$ calculated neglecting the effect of the atmosphere -- 
remains valid as soon as the mass of the H$_2$O-dominated atmosphere is significantly lower than the total mass 
of the planet.

Lastly, we drew mass-radius relationships for rocky planets with various water-to-rock mass ratios. This last step 
was performed by carrying out a logarithmic interpolation of the water-to-rock mass ratio for each possible rocky core mass and radius using 
the array of transit radius calculated for a wide range of possible total H$_2$O atmospheric pressures\footnote{The mass of the 
H$_2$O-dominated atmosphere is calculated by summing the mass of each atmospheric layer in our 1-D radiative-convective model.}

The second step of our procedure (i.e., the calculation of z$_{\text{atmosphere}}$) was achieved through a number of substeps listed below:
Firstly, we estimated the surface temperature $\text{T}_\text{surf}$ of a H$_2$O-dominated steam atmosphere 
as a function of H$_2$O atmospheric pressure P$_{\text{H}_2\text{O}}$, surface gravity $g$ and 
irradiation received by the planet $S_{\text{eff}}$, following the same approach as in \citealt{Turbet:2019aa} (Fig.~3). 
To do this, we first performed 1-D inverse radiative-convective calculations for a wide range of surface temperatures 
(from 300 to 4300~K), irradiations (roughly from 1 to 40$\times$ the irradiation received on Earth), surface gravities 
(from 2 to 50~m~s$^{-2}$), and water vapor pressures (from 2.7$\times$10$^5$ to 2.7$\times$10$^9$~Pa). 
We then fit a polynomial (see Methods in Appendix~\ref{appendix_polynomial_fit}) on all these parameters to derive the following 
empirical equation for the surface temperature $\text{T}_\text{surf}$ (in Kelvins): 
\begin{equation}
\begin{split}
& \log_{10}{\Big(\text{T}_\text{surf}}(x,y,z)\Big)~=~c_1+c_2~x+c_3~y+c_4~z \\
&+c_5~x^2+c_6~x~y+c_7~y^2+c_8~z^2+c_9~y^3+c_{10}~z^3, \\ 
\end{split}
\label{empirical_equation_tsurf}
\end{equation}
with $x$~=~($\log_{10}{(P_{\text{H}_2\text{O}})}$~-~k$_1$)/k$_2$ with $P_{\text{H}_2\text{O}}$ the H$_2$O partial pressure expressed in bar units, 
$y$~=~($\log_{10}{(g)}$~-~k$_3$)/k$_4$ with $g$ the surface gravity (at the interior-atmosphere boundary) in m~s$^{-2}$, 
and $z$~=~($\log_{10}{(S_{\text{eff}})}$~-~k$_5$)/k$_6$ with $S_{\text{eff}}$ 
the irradiation received by the planet (S$_{\text{eff}}$ is in Earth insolation units; i.e., $S_{\text{eff}}$~=~1 when the planet 
receives the same insolation as Earth of 1366~W~m$^2$). 
The empirical coefficients are shown in Table~\ref{tab:table_coeff_tsurf}.

\begin{table}[h!]
  \begin{center}
    \caption{Coefficients for the polynomial fit of the surface temperature of a H$_2$O steam atmosphere, 
presented in Equation~\ref{empirical_equation_tsurf}.}    \label{tab:table_coeff_tsurf}
    \begin{tabular}{l|c||c|r}
      \textbf{Coefficient} & \textbf{Value} & \textbf{Coefficient} & \textbf{Value} \\
 & & & \\
k$_1$ & 2.688 & k$_{4}$ & 4.683~$\times$10$^{-1}$ \\
k$_2$ & 1.019 & k$_{5}$ & 7.664~$\times$10$^{-1}$ \\
k$_3$ & 1.099 & k$_{6}$ & 4.224~$\times$10$^{-1}$ \\
 & & & \\
c$_1$ & 3.401 & c$_{6}$ & 8.519~$\times$10$^{-3}$ \\
c$_2$ & 1.501~$\times$10$^{-1}$ & c$_{7}$ & -1.467~$\times$10$^{-2}$ \\
c$_3$ & -3.146~$\times$10$^{-2}$ & c$_{8}$ & -7.091~$\times$10$^{-3}$ \\
c$_4$ & 4.702~$\times$10$^{-2}$ & c$_{9}$ & -7.627~$\times$10$^{-3}$ \\
c$_5$ & -4.911~$\times$10$^{-3}$ & c$_{10}$ & 8.348~$\times$10$^{-3}$ \\
    \end{tabular}
  \end{center}
\end{table}

This empirical relationship provides an estimate 
of the surface temperature (see Fig~\ref{tsurf_contourplot}) of a H$_2$O-dominated steam atmosphere as a function 
of surface gravity, water vapor surface pressure and irradiation. 
It is valid within a few percent for most of the parameter space (maximum error $\sim$~10$\%$; see Fig.~\ref{figure_fit2}, left panel), 
for irradiation from $\sim$~1 to 30~S$_\oplus$ (assuming the irradiation received by the planet is above 
the runaway greenhouse irradiation threshold), surface gravity from 0.2 to 6~g$_\oplus$, 
and water vapor pressure from 2.7~bar to 27~kbar, and as far as 
surface temperature remains between 300 and 4300~K.

Secondly, for each possible rocky core mass and radius pair taken from \citet{Zeng:2016}, and for a wide range of 
H$_2$O atmospheric pressures (from 2.7$\times$10$^{1}$ to 2.7$\times$10$^{5}$~bars), we built the atmospheric structure following 
the approach presented in \citealt{Turbet:2019aa} (Appendix~\ref{appendix_trick_r-m}), 
and originating from \citealt{Marcq:2012}, \citealt{Marcq:2017} and \citealt{Pluriel:2019}.

Lastly, we evaluated the transit radius of each possible planet 
(made of each possible combination of rocky interior and water atmospheric pressure) 
by integrating these atmospheric profiles in the hydrostatic approximation, 
using nonideal thermodynamic properties of H$_2$O \citep{Haar:1984} as in \citealt{Turbet:2019aa}, 
and assuming the transit radius is controlled by the altitude of the upper water cloud layer. 
For this, we used the altitude of the top of the moist convective layer as a proxy. 
The total atmospheric transit thickness of a thick H$_2$O-dominated atmosphere has been shown to be roughly unchanged whether 
a cloudy or cloud-free atmosphere is considered \citep{Turbet:2019aa}.

\subsection{An empirical mass-radius relationship formula}

Motivated to make our revised mass-radius relationships accessible to the community, 
we constructed an empirical mass-radius relationship formula (provided below) for water-rich rocky planets receiving more irradiation 
than the runaway greenhouse irradiation limit. This formula was constructed in two steps:

Firstly, we derived an analytic expression of the mass-radius relationships, 
assuming (i) the perfect gas law approximation, and (ii) an isothermal temperature profile:
\begin{equation}
z_{\text{atmosphere}}~=~\Big(\frac{1}{\log{\big(\frac{\text{x}_{\text{H}_2\text{O}}}{1-\text{x}_{\text{H}_2\text{O}}} \times \frac{\text{g}_{\text{core}}^2}{4 \pi~\text{G}~\text{P}_{\text{transit}}}\big)} \times \frac{\text{R}~\text{T}_{\text{eff}}}{\text{M}_{\text{H}_2\text{O}}~\text{g}_{\text{core}}}}-\frac{1}{~\text{R}_{\text{core}}~}\Big)^{-1,}
\label{empirical_equation_thickness}
\end{equation}
with R$_{\text{core}}$ and g$_{\text{core}}$ the core (or surface) radius and gravity of the planet, respectively,
R the gas constant (=~8.314~J~K$^{-1}$~mol$^{-1}$), M$_{\text{H}_2\text{O}}$ the molar mass of water (=~1.8$\times$10$^{-2}$~kg~mol$^{-1}$), 
G the gravitational constant (=~6.67$\times$10$^{-11}$~m$^3$~kg$^{-1}$~s$^{-2}$), 
and x$_{\text{H}_2\text{O}}$ the water mass fraction (between 0 and 1) of the planet. 
P$_{\text{transit}}$ is the pressure at the transit radius. T$_{\text{eff}}$ is the temperature of the isothermal atmosphere.
The procedure to derive this equation is detailed in Appendix~\ref{appendix_fit_thickness}.
This equation well describes (see hereafter) the family of possible behaviors 
of the mass-radius relationships for H$_2$O steam atmosphere planets.

Secondly, we fit the free parameters (T$_{\text{eff}}$ and P$_{\text{transit}}$) using 
the range of simulations described in the previous subsection. Our simulations show that P$_{\text{transit}}$ varies little 
across the range of parameters we explored and is roughly equal to 10$^{-1}$~Pa. We thus set it to this value.
T$_{\text{eff}}$ is an effective atmospheric temperature that we empirically fit 
(see Methods in Appendix~\ref{appendix_polynomial_fit}) as follows:
\begin{equation}
\begin{split}
& \log_{10}{\Big(\text{T}_\text{eff}}(x,y,z)\Big)~=~{\beta}_1+{\beta}_2~x+{\beta}_3~y+{\beta}_4~z \\
&+{\beta}_5~x~y+{\beta}_6~y^2+{\beta}_7~x^3+{\beta}_8~x^2~y+{\beta}_9~x~y^2+{\beta}_{10}~y^4, \\ 
\end{split}
\label{empirical_equation_teff}
\end{equation}
with $x$~=~($\log_{10}{(\text{x}_{\text{H}_2\text{O}})}$~-~${\alpha}_1$)/${\alpha}_2,$ 
with $\text{x}_{\text{H}_2\text{O}}$ the mass water fraction of the planet (between 0 and 1), 
$y$~=~($\log_{10}{(g)}$~-~${\alpha}_3$)/${\alpha}_4,$ with $g$ the surface gravity (at the interior-atmosphere boundary) in m~s$^{-2}$, 
and $z$~=~($\log_{10}{(S_{\text{eff}})}$~-~${\alpha}_5$)/${\alpha}_6,$ with $S_{\text{eff}}$ 
the irradiation received by the planet (S$_{\text{eff}}$ is in Earth insolation units). 
The empirical coefficients are shown in Table~\ref{tab:table_coeff_teff}.

These relationships (Equations~\ref{empirical_equation_thickness}
 and \ref{empirical_equation_teff}) are valid 
within a few percent for most of the parameter space (again, maximum error of $\sim$~10$\%$; 
see Fig.~\ref{figure_fit2}, right panel), for irradiation from 1 to 30~S$_\oplus$ 
(assuming the irradiation received by the planet is above 
the runaway greenhouse irradiation threshold), surface gravity from 0.2 to 6~g$_\oplus$, 
and water vapor pressure from 2.7~bar to 27~kbar, and as far as 
surface temperature remains between 300 and 4300~K. 
\begin{table}[h!]
  \begin{center}
    \caption{Coefficients for the polynomial fit of the H$_2$O steam atmosphere 
effective temperature presented in equation~\ref{empirical_equation_teff}.}    \label{tab:table_coeff_teff}
    \begin{tabular}{l|c||c|r}
      \textbf{Coefficient} & \textbf{Value} & \textbf{Coefficient} & \textbf{Value} \\
 & & & \\
${\alpha}_1$ & -3.550 & ${\alpha}_{4}$ & 4.683~$\times$10$^{-1}$ \\
${\alpha}_2$ & 1.310 & ${\alpha}_{5}$ & 7.664~$\times$10$^{-1}$ \\
${\alpha}_3$ & 1.099 & ${\alpha}_{6}$ & 4.224~$\times$10$^{-1}$ \\
 & & & \\
${\beta}_1$ & 2.846 & ${\beta}_{6}$ & 1.736~$\times$10$^{-2}$ \\
${\beta}_2$ & 1.555~$\times$10$^{-1}$ & ${\beta}_{7}$ & 1.859~$\times$10$^{-2}$ \\
${\beta}_3$ & 8.777~$\times$10$^{-2}$ & ${\beta}_{8}$ & 4.314~$\times$10$^{-2}$ \\
${\beta}_4$ & 6.045~$\times$10$^{-2}$ & ${\beta}_{9}$ & 3.393~$\times$10$^{-2}$ \\
${\beta}_5$ & 1.143~$\times$10$^{-2}$ & ${\beta}_{10}$ & -1.034~$\times$10$^{-2}$ \\
    \end{tabular}
  \end{center}
\end{table}

We propose in Appendix~\ref{appendix_how_to} a tutorial on how to use these mass-radius relationships.

\section{Results}
\label{section_results}

\subsection{Revised mass-radius relationships}
\begin{figure*}
    \centering
\includegraphics[width=\linewidth]{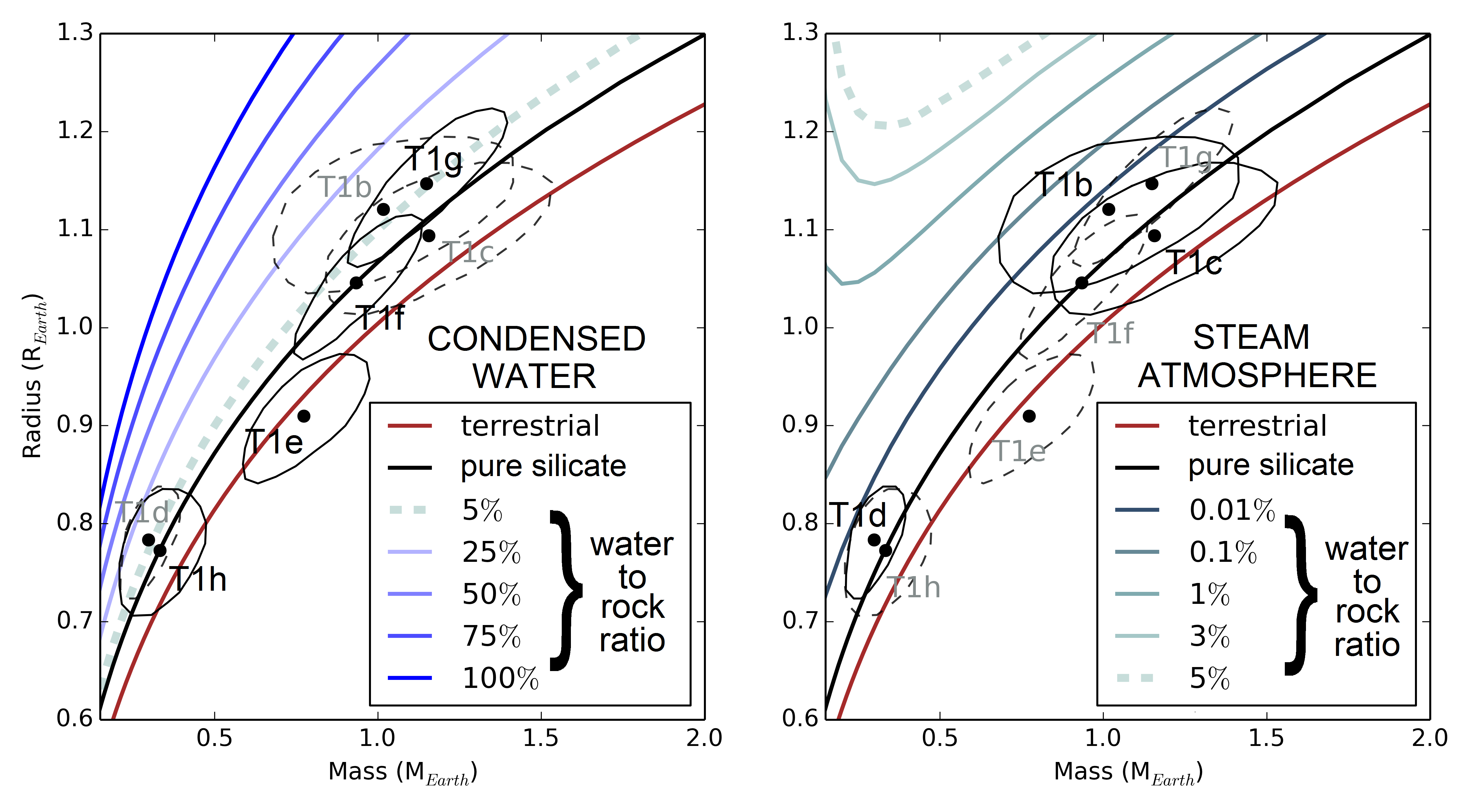}
\caption{Mass-radius relationships for various interior compositions and water content, assuming water is in the condensed 
form (left panel) and water forms an atmosphere (right panel). The silicate composition 
mass-radius relationship assumes a pure MgSiO$_3$ interior and 
was taken from \citet{Zeng:2016}. The water-rich mass-radius 
relationships for water in condensed form (left panel) were derived using the data from \citet{Zeng:2016}. 
The water-rich mass-radius relationships for water in gaseous form (right panel) are the result of the present work. 
All mass-radius relationships with water were built assuming a pure MgSiO$_3$ interior. 
For comparison, we added the measured positions of the seven 
TRAPPIST-1 planets measured from \citet{Grimm:2018}, with their associated 95$\%$ confidence ellipses. 
Based on the irradiation they receive compared to the theoretical runaway greenhouse limit \citep{Kopparapu:2013,Wolf:2017,Turbet:2018aa}, 
TRAPPIST-1 e, f, g, and h should be compared with mass-radius relationships on the left, while TRAPPIST-1b, c, and d should be compared with those on the right. 
To emphasize this, we indicated, on each panel and in black (and solid line ellipses), the planets (and their associated 
95$\%$ confidence ellipses) for which mass-radius relationships (with water) are appropriate. 
In contrast, we indicated on each panel in gray (and dashed line ellipses) the planets (and their associated 
95$\%$ confidence ellipses) for which mass-radius relationships (with water) are not appropriate. 
For reference, we also added a terrestrial composition that resembles that of the Earth, but also that of Mars and Venus. 
We note that mass-radius relationships for 
steam planets (right panel) can be easily built following the procedure described in Appendix~\ref{appendix_how_to}.}
\label{m-r_relationship}
\end{figure*}

The main result of this work is summarized in Figure~\ref{m-r_relationship}, which shows how mass-radius relationships 
can vary depending on if water is treated as a condensed layer \citep{Zeng:2016} or as an atmosphere (this work).
As a direct consequence of the runaway greenhouse radius inflation introduced in \citet{Turbet:2019aa}, 
mass-radius relationships in the steam atmosphere configuration give -for a given planet mass- 
a significantly larger radius than in the condensed water configuration. This translates in two main consequences:
Firstly, traditional mass-radius relationships \citep{Seager:2007,Sotin:2007,Grasset:2009,Mordasini:2012,Swift:2012,Zeng:2013,Zeng:2016} 
for water-rich rocky planets (i.e. where most water is considered to be in the solid or liquid form) tend to significantly 
overestimate their bulk density if the planets are more irradiated than the runaway greenhouse irradiation limit. 
Secondly, comparing these traditional mass-radius relationships for water-rich rocky planets 
with real planet measured densities tend to overestimate 
the evaluation of their water-to-rock mass fraction, possibly by several orders of magnitude. 

 In Equations~\ref{empirical_equation_thickness} and \ref{empirical_equation_teff}, we provide
an empirical formula for the H$_2$O steam atmosphere thickness as a function of 
planet core gravity and radius, water content and irradiation. This formula can easily be used 
(see the procedure in Appendix~\ref{appendix_how_to}) 
to construct mass-radius relationships for 
water-rich, rocky planets that are more irradiated than the runaway greenhouse irradiation threshold, for any type of planet interior.

Lastly, our revised mass-radius relationships for steam planets indicate 
that small rocky planets (M$_{\text{planet}}$~$\lessapprox$~0.5M$_{\oplus}$) that are more irradiated than the 
runaway greenhouse irradiation threshold should be unable to retain more than a few percent water by mass. 
This is because for these small planets the runaway-greenhouse-induced radius inflation is so extreme that 
the upper atmosphere becomes gravitationally unbounded for steam atmospheres -only a few percent by mass- 
and efficient atmospheric escape mechanisms should take place. 
For instance, for a 0.3M$_{\oplus}$ pure silicate core planet (located just above 
the runaway greenhouse irradiation threshold) with a 5$\%$ water-to-rock ratio, Figure~\ref{m-r_relationship} (right panel) indicates 
that the transit radius lies around 1.2R$_{\oplus}$. The gravity at the transit 
radius is thus as low as 20$\%$ of that at the 
surface of the Earth, so $\sim$~2~m~s$^{-2}$, meaning atmospheric escape can be very strong. 
In fact, the U-shape of the mass-radius relationships (in the upper-left part 
of the mass-radius relationships for steam planets in Figure~\ref{m-r_relationship}) is symptomatic of the fact that 
the atmosphere becomes gravitationally unbounded. This U-shape has already been predicted 
for H$_2$/He-rich planets \citep{Fortney:2007,Baraffe:2008,Lopez:2014,Zeng:2019}, 
but we show here that it is also expected for planets endowed with H$_2$O-rich atmospheres.
This U-shape can be described well at first order by 
Equation~\ref{empirical_equation_thickness}.

\subsection{Application to the TRAPPIST-1 system}

The fact that the use of traditional mass-radius relationships for water-rich rocky planets 
tend to overestimate the evaluation of a planet water-to-rock mass fraction 
is particularly relevant for our understanding of the nature of the TRAPPIST-1 planets, 
the only system known to date (as of November 2019) of temperate-orbit Earth-size planets \citep{Gillon:2017} 
for which both radii and masses have been measured \citep{Grimm:2018}. 
Based on comparisons of TRAPPIST-1 planet bulk densities with traditional mass-radius relationships, it has been 
speculated that some planets in the system may be enriched with water, 
possibly up to tens of percent for some of them \citep{Grimm:2018,Dorn:2018}.

\begin{table}[h!]
  \begin{center}
    \caption{Maximum water content of TRAPPIST-1 b, c, and d, depending on the assumed core composition. 
Maximum water mass fractions were obtained by finding the corresponding mass-radius relationships 
that pass just above the 95$\%$ confidence ellipses from \citet{Grimm:2018}.}    \label{tab:max_water_T1}
    \begin{tabular}{c|c|c|c}
      \textbf{Core composition} & \multicolumn{3}{c}{\textbf{Maximum H$_2$O mass fraction}} \\
       & T-1b & T-1c & T-1d \\
      \hline
      pure silicate (MgSiO$_3$) & 0.4$\%$ & 0.01$\%$ & 0.01$\%$ \\
      terrestrial & 2$\%$ & 0.3$\%$ & 0.08$\%$ \\
      pure iron & $>$10$\%$ & $>$10$\%$ & 2$\%$ \\
    \end{tabular}
  \end{center}
\end{table}

Our results suggest that the three innermost planets of the TRAPPIST-1 system -and more particularly, TRAPPIST-1 b and d, for which 
TTVs measurements point toward particularly low bulk densities \citep{Grimm:2018}- 
do not necessarily need to be highly enriched with water to reach their measured density. 
In fact, Table~\ref{tab:max_water_T1} provides 
quantitative estimates for the maximum water content of the three TRAPPIST-1 innermost planets, for several core compositions. 
For a core composition similar to that of the solar system terrestrial planets, 
TRAPPIST-1 b, c, and d cannot accommodate more than 2, 0.3, and 0.08$\%$, respectively, of water. 
Specifically, TRAPPIST-1 d cannot be composed of more than 2$\%$ water whatever the core composition assumed.
For comparison, \citet{Bourrier:2017b} evaluated that the current rate of water loss can be as high as 
0.19, 0.06, and 0.18$\%$ per gigayear by mass for TRAPPIST-1 b, c, and d, respectively. 
Putting these pieces of information together, it is likely that the three inner TRAPPIST-1 planets 
may all be completely dry today.

A direct consequence of this result is that if the planets of the TRAPPIST-1 system are all rich in water, 
as supported by planet formation and migration models for which TRAPPIST-1 planets formed far from 
their host star, beyond water and other volatile ice lines, 
and subsequently migrated forming a resonant chain \citep{Ormel:2017,Unterborn:2018,Coleman:2019}, 
then our revised mass-radius relationships -leading to much lower water content for TRAPPIST-1 
inner planets than previous calculations \citep{Grimm:2018,Dorn:2018} showed- can be reconciled 
with the fact that outer planets are expected to be more volatile-rich and water-rich than inner planets \citep{Unterborn:2018}, 
due both to planet formation and migration \citep{Ormel:2017, Unterborn:2018,Coleman:2019}, 
and atmospheric escape processes \citep{Bolmont:2017,Bourrier:2017b}.
This would avoid the need for exotic planet formation and water delivery processes \citep{Dorn:2018,Schoonenberg:2019} to explain 
apparent density variation with irradiation among TRAPPIST-1 planets.

\medskip

As of November 2019, the uncertainties on the masses of TRAPPIST-1 planets are still large 
(see the 2-$\sigma$~uncertainty ellipses on Fig.~\ref{m-r_relationship}, from \citealt{Grimm:2018}). 
However, it is expected that these uncertainties will significantly decrease in the near future, either 
through a follow-up on the transit timing variations 
(Spitzer Proposal ID 14223, PI: Eric Agol) or using radial velocity measurements with near-infrared 
ground-based spectrographs \citep{Klein:2019} such as SPIRou \citep{Artigau:2014} or NIRPS \citep{Wildi:2017}.
\begin{figure*}
    \centering
\includegraphics[width=18cm]{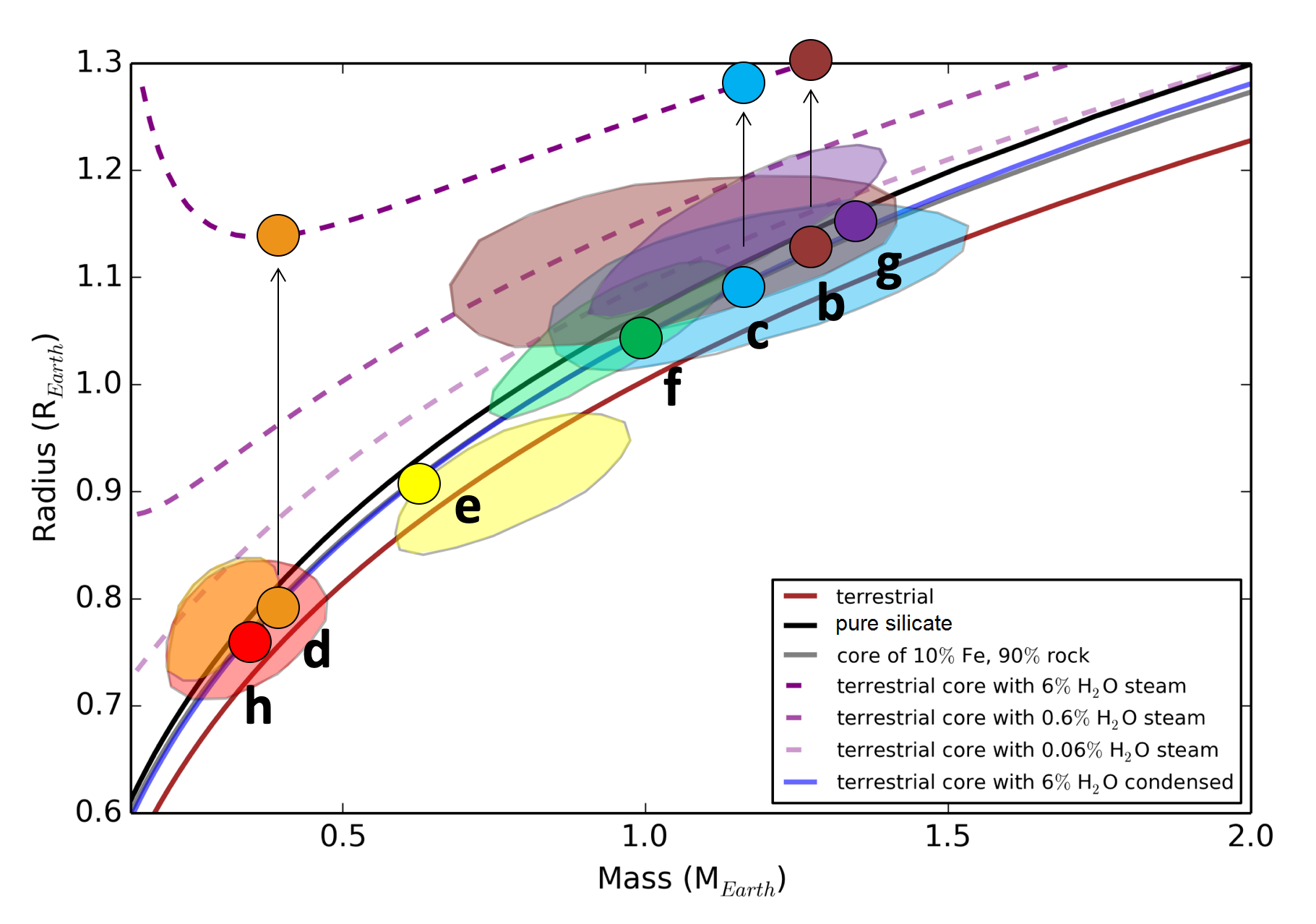}
\caption{Example of a scenario for TRAPPIST-1 planets where masses and radii 
follow an interior isocomposition line (gray line; 10$\%$ Fe, 90$\%$ MgSiO$_3$ composition) chosen 
to be consistent with 2-$\sigma$~uncertainty ellipses of \citealt{Grimm:2018}. 
While the seven large ellipses indicate the known current estimates (95$\%$ confidence) for the 
masses and radii of the seven TRAPPIST-1 planets, 
the seven small circles indicate the positions (in the mass, radius diagram) of the seven planets as speculated in our scenario. 
Each planet (associated with a current uncertainty ellipse, and a speculated position) is identified by a distinct color. 
This figure also shows mass-radius relationships for terrestrial core planets, in some cases endowed with either a condensed layer 
of water (solid blue line) or a steam H$_2$O atmosphere of various masses (dashed purple lines). The mass-radius relationships for 
steam planets can be built following the procedure described in Appendix~\ref{appendix_how_to}.
We note that the mass-radius relationships for steam H$_2$O-rich atmosphere planets can slightly change depending 
on the level of stellar irradiation they receive. However, because these changes are low (for the levels of irradiation on TRAPPIST-1 b, 
c, and d) compared to the runaway-greenhouse-transition-induced mass-radius relationship change, we decided no to show them for clarity.}
\label{scenarios_density_trappist-1}
\end{figure*}

Below, and with the support of Fig.~\ref{scenarios_density_trappist-1}, 
we discuss, as a proof of concept, one example of a possible scenario for the masses and radii of each of the seven TRAPPIST-1 planets, 
that of the case where all the TRAPPIST-1 planets closely follow an isocomposition interior mass-radius relationship. 
The baseline interior composition (10$\%$ Fe, 90$\%$ MgSiO$_3$, i.e., the solid gray line in Fig.~\ref{scenarios_density_trappist-1}) 
was chosen to ensure that this scenario remains 
compatible with the \citet{Grimm:2018} 95$\%$ confidence ellipses\footnote{Note that this condition 
requires having a core composition of at least $\sim$~90$\%$ silicate (MgSiO$_3$).}. 

This assumption, however, does not guarantee in principle that the planets do have an interior composition of 10$\%$ Fe and 90$\%$ MgSiO$_3$. 
This stems from the fact that the composition of the planets is, in principle, highly degenerate, because 
their positions in the mass-radius diagram (Fig.~\ref{scenarios_density_trappist-1}) can be explained 
either by (i) dry planets with a 10$\%$ Fe~+~90$\%$ MgSiO$_3$ core (solid gray line in Fig.~\ref{scenarios_density_trappist-1}), 
or (ii) wet planets with a denser core (e.g., 6$\%$ water with a terrestrial core, as illustrated by 
the solid blue line in Fig.~\ref{scenarios_density_trappist-1}). This is illustrated in Fig.~\ref{scenarios_density_trappist-1} where 
the solid blue (6$\%$ water with a terrestrial core) and gray (10$\%$ Fe~+~90$\%$ MgSiO$_3$ core) lines are almost superimposed.

However, this degeneracy is removed here bearing in mind that the three innermost planets of the TRAPPIST-1 system are 
more irradiated than the runaway greenhouse limit and should therefore follow a different isocomposition mass-radius relationship 
(e.g., the upper dashed purple line in Fig.~\ref{scenarios_density_trappist-1}, for planets
with 6$\%$ water and a terrestrial core). The black arrows in Fig.~\ref{scenarios_density_trappist-1} indicate 
the new positions of TRAPPIST-1 b, c, and d in the mass-radius diagram taking into account the revised mass-radius relationship.
In other words, the runaway greenhouse transition allow planets to jump from one 
mass-radius relationship to another, which makes it possible to break the composition degeneracy.

This demonstrates, to a certain extent, that in our scenario all the TRAPPIST-1 planets should all be very dry, because 
(i) if all planets were to be water-rich, then they would have to 
follow a different mass-radius relationship (purple dashed lines for the three innermost planets, 
versus solid blue line for the four outermost planets in Fig.~\ref{scenarios_density_trappist-1}); 
(ii) if only some of the planets were to be water-rich and others were not, 
then the planets should not follow an isocomposition mass-radius relationship anyway. 
In our scenario, we evaluate that, assuming that all TRAPPIST-1 planets have the same mass composition 
(for the rocky interior and water content), the planets cannot accommodate more than 10$^{-3}\%$ of water by mass
in order to fit all the small circles in Fig.~\ref{scenarios_density_trappist-1}.
This argument -that all planets are very dry- should hold unless we are dealing with a fine-tuned scenario where, for each of the planets, 
all processes (water delivery, runaway greenhouse radius inflation effect, water loss, different core composition) 
compensate each other exactly.

A direct consequence is that any significant deviation of planetary densities from an interior isocomposition 
mass-radius relationship would be a strong indication that (i) there is either today large reservoirs of water or volatiles
on at least some planets of the system, or that (ii) there are significant differences in TRAPPIST-1 planets' core composition. 
In some cases (e.g., a significant trend in planets density with irradiation), the first interpretation would 
be favored.

\section{Conclusions}
\label{section_conclusions_discussions}

In this work, we calculated revised mass-radius relationships for water-rich, rocky planets, 
which are more irradiated than the runaway 
greenhouse irradiation limit. This was performed by coupling the 
mass-radius relationships for rocky interior of \citet{Zeng:2016} with our 
estimates of the atmospheric thickness of H$_2$O-dominated atmospheres with a 1-D radiative-convective model. 

For a given water-to-rock mass ratio, our revised mass-radius relationships lead to planet bulk densities much lower 
than calculated when most water is assumed to be in condensed form, which is the common standard in the literature 
\citep{Seager:2007,Sotin:2007,Grasset:2009,Mordasini:2012,Swift:2012,Zeng:2013,Zeng:2016}. This means that using 
traditional mass-radius relationships for planets that are more irradiated than the 
runaway greenhouse irradiation limit tends to dramatically overestimate -possibly by several orders of magnitude- 
their bulk water content. 

More specifically, this result has important consequences 
for our understanding of the nature of the TRAPPIST-1 planets. Our work shows that the 
measured density (yet to be confirmed)- of the three innermost planets 
of the TRAPPIST-1 system indicates their bulk water content should be significantly lower than what was previously speculated 
in \citet{Grimm:2018} and \citet{Dorn:2018}.
More generally, these results demonstrate that non-H$_2$/He-dominated atmospheres can have a first-order effect on 
the mass-radius relationships even for Earth-mass planets receiving moderate irradiation.

Future work should focus on more carefully taking into account possible interactions and feedback between the planet interior 
and the steam atmosphere, and should aim to extend our work to more irradiated, more massive planets (so-called super-Earth planets), for 
which mass and radius measurements have been performed for a much larger number of planets. 
For this, an interior model could be coupled to a steam atmosphere model to account for (1) the greenhouse effect feedback of the 
atmosphere on the interior structure; (2) the planetary core cooling; and (3) the possible outgassing or accumulation 
through photodissociation of various gases such as O$_2$, N$_2$, CO$_2$, etc. 
Future work should also re-examine our results with 3-D global climate models, 
consistently taking into account the effect of clouds and short-wave absorption in the upper atmosphere. 
This is in order to improve the estimate of the thermal structure and thus the true radius of the planet.
Meanwhile, in 
Section~\ref{section_methods} we provide empirical formulae for the surface temperature and the thickness of a H$_2$O steam atmosphere, as well as 
a tutorial on how to correctly use them in Appendix~\ref{appendix_how_to}. 
These formulae can be used in interior models to better capture the boundary effect of a 
thick H$_2$O-dominated atmosphere.

\begin{acknowledgements}
This project has received funding from the European Union’s Horizon 2020 research and 
innovation program under the Marie Sklodowska-Curie Grant Agreement No. 832738/ESCAPE. 
This project has received funding from the European Research Council (ERC) under 
the European Union’s Horizon 2020 research and innovation program (grant agreement No. 724427/FOUR ACES 
and No. 679030/WHIPLASH). 
This work has been carried out within the framework of the National Centre of Competence in Research PlanetS 
supported by the Swiss National Science Foundation. The authors acknowledge the financial support of the SNSF.
M.T. is grateful for the computing resources on OCCIGEN (CINES, French National HPC).
This research has made use of NASA's Astrophysics Data System.
M.T. thanks the Gruber Foundation for its generous support to this research. 
M.T. thanks Brice-Olivier Demory for providing useful feedbacks on the manuscript.
Last but not least, we thank the reviewer for his/her insightful remarks and comments on our manuscript.
\end{acknowledgements}

\bibliographystyle{aa} 
\interlinepenalty=10000
\bibliography{biblio} 

\appendix 

\section{Why and when the R$_{\text{planet}}$~=~R$_{\text{core}}$+z$_{\text{atmosphere}}$ approximation is valid}
\label{appendix_trick_r-m}

In order to calculate mass-radius relationships for water-rich rocky planets, we assumed that the transit radius of a planet 
can be approximated by the sum of the core radius (directly taken from the \citealt{Zeng:2016} dry mass-radius relationships) and the 
thickness of the water layer, calculated independently. 
This approach remains valid only if the feedback of the water layer on the rocky interior physical size is negligible. 
The presence of a water layer (in solid, liquid, or gaseous form) can have two distinct impacts:

\begin{figure}
    \centering
\includegraphics[width=\linewidth]{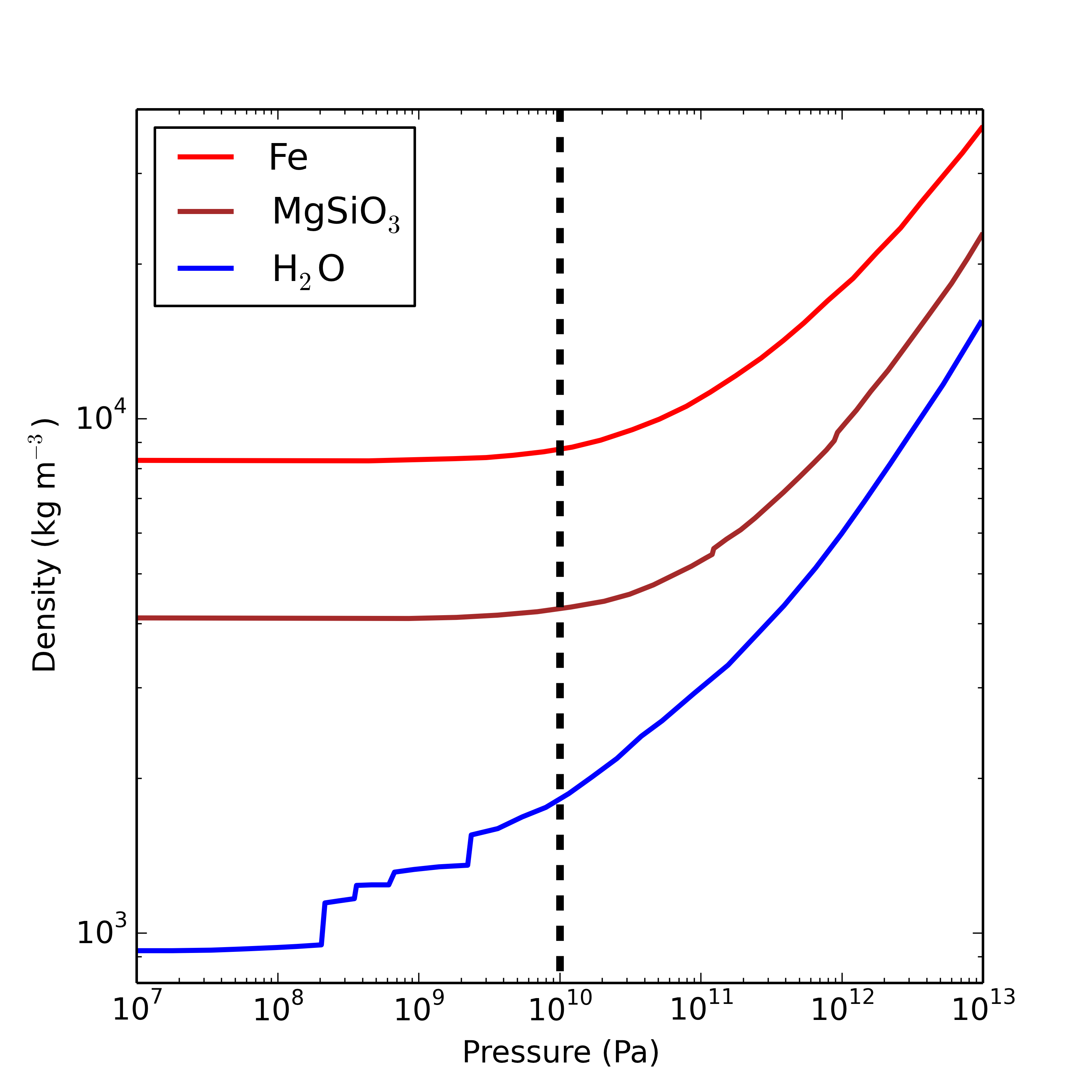}
\caption{Equations of state (EOS) for iron, 
silicate (MgSiO$_3$; perovskite phase and its high-pressure derivatives), and  H$_2$O (Ice Ih, Ice III, Ice V, Ice 
VI, Ice VII, Ice X, and superionic phase along its melting curve, i.e., solid-liquid phase boundary). These EOS 
were taken from \citealt{Zeng:2013} (Figure~1). The vertical, dashed line denotes the typical pressure at which 
the density of iron and MgSiO$_3$ starts to deviate from a constant value.}
\label{EOS_zeng}
\end{figure}

\begin{figure}
    \centering
\includegraphics[width=\linewidth]{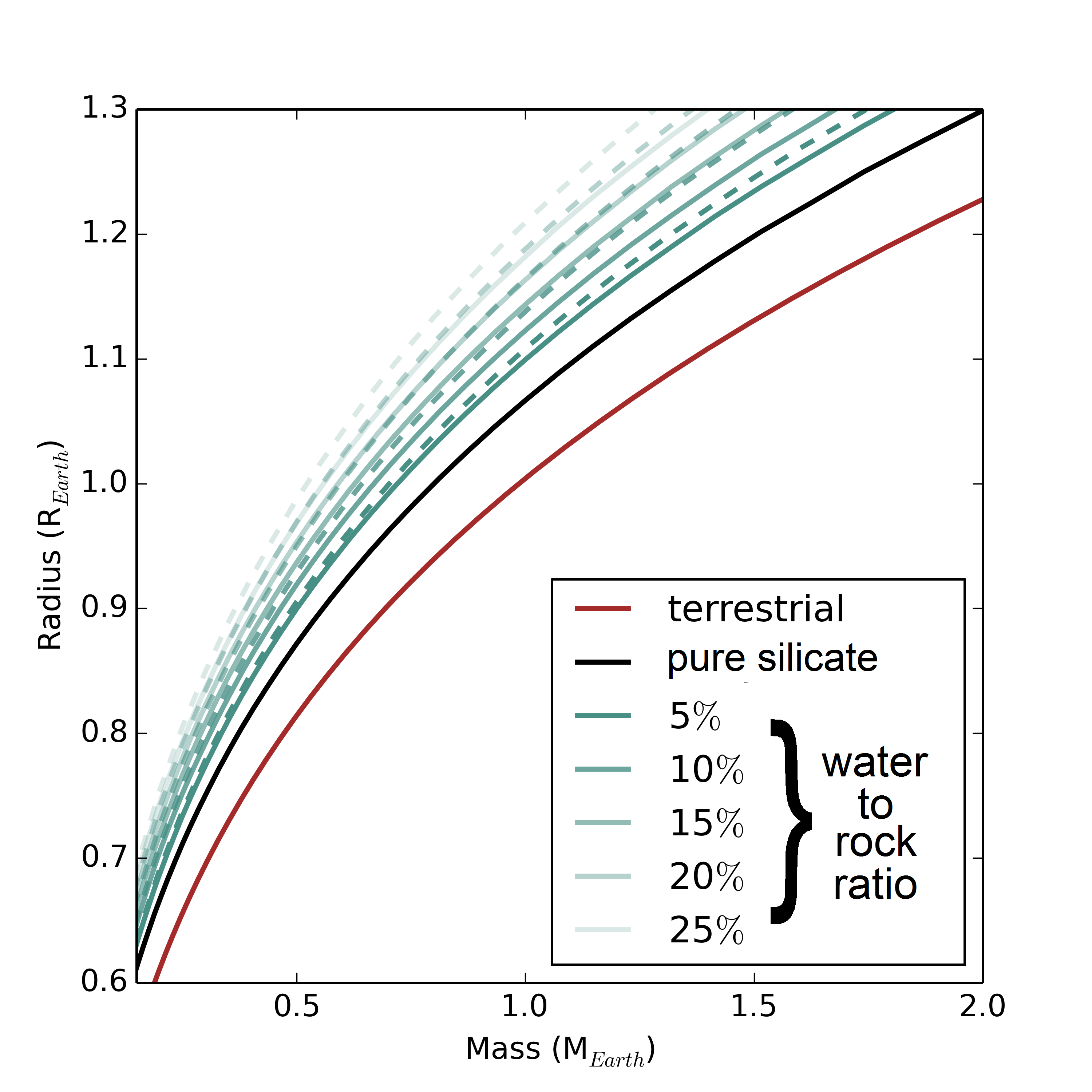}
\caption{Comparisons of mass-radius relationships for water-rich planets 
(with water in condensed phase) of \citet{Zeng:2016} (solid blue lines) with mass-radius relationships calculated 
in the present work (dashed blue lines), and assuming a layer of condensed water is added on top of mass-radius relationships 
for pure silicate (MgSiO$_3$) planets of \citet{Zeng:2016}.}
\label{compare_zeng}
\end{figure}

\begin{enumerate}
 \item The water layer can exert a pressure force that compresses the rocky interior. Figure~\ref{EOS_zeng} shows 
the equation of state (EOS) for silicate (MgSiO$_3$, denoted by a solid brown line). The density of MgSiO$_3$ is roughly constant 
until pressure reaches $\sim$~10$^{10}$~Pascals (denoted by the vertical, dashed line). 
In other words, it means that if the water layer exerts a basal pressure that is significantly lower than this $\sim$~10$^{10}$~Pascals limit, 
then the presence of the water layer should have a negligible effect on the silicate interior density profile and thus its physical size. 
To check that this effect does not significantly affect the mass-radius relationships presented in Figure~\ref{m-r_relationship}, 
we calculated mass-radius relationships of water-rich planets with a pure silicate interior, 
with water assumed to be present in a solid layer (using the water EOS shown in Figure~\ref{EOS_zeng}, denoted by a solid blue line), 
and compared these calculations with those of \citet{Zeng:2016} that self-consistently take into account the pressure feedback on the interior. 
The result of this comparison is shown in Figure~\ref{compare_zeng}. For a 5$\%$ water-to-rock mass ratio, 
the approximation made in our work leads to a 1$\%$~error maximum (for a 2~M$_{\oplus}$ core planet) for the range 
of planets discussed in Figures~\ref{m-r_relationship} and \ref{compare_zeng}. 
Finally, this demonstrates that the approximation discussed here is largely acceptable to establish the 
mass-radius relationships presented in Figure~\ref{m-r_relationship}. 

 \item The water layer can change the thermal structure and possibly even the physical state of the interior. 
This is particularly relevant in the H$_2$O steam atmosphere case where 
the surface temperature can reach thousands of Kelvin \citep{Kopparapu:2013,Goldblatt:2013,Turbet:2019aa}, 
which imposes an extreme surface boundary condition on the interior. 
While the direct interior temperature profile change should have a limited impact on the radius of the rocky core \citep{Seager:2007,Zeng:2014}, 
the temperature change could lead to a phase change of the interior (e.g., melting) that could significantly increase its physical 
radius \citep{Bower:2019}. Taking this effect into account in a self-consistent way 
requires the use of an interior atmosphere coupled model. We leave this for future work. 
\end{enumerate}

\section{Procedure for the polynomial fits}
\label{appendix_polynomial_fit}

The polynomial fit of our data (surface temperature and effective temperature) was performed in four distinct steps and makes use of 
the scikit-learn python library \citep{Scikit:2011}.

As a first step, we recentered and normalized the distribution of values for 
each parameter (water pressure or water content, surface gravity, stellar flux) of the fit. 
For this, we used the \textit{StandardScaler} python tool\footnote{\url{https://scikit-learn.org/stable/modules/generated/sklearn.preprocessing.StandardScaler.html}}.
As a second step, we built a matrix of all possible terms of polynomials of degree $n$ or lower. This matrix was constructed using the 
\textit{PolynomialFeatures} python tool\footnote{\url{https://scikit-learn.org/stable/modules/generated/sklearn.preprocessing.PolynomialFeatures.html}}. 
As a reminder, for a polynomial of degree $n$ constructed on $k$ parameters, there is a total number of 
$N$~=~$\dbinom{k+n}{n}$~=~$\frac{(k+n)!}{n!~k!}$ polynomial terms. 
In practice, we constructed a matrix of all possible terms 
of polynomials of degree $n$~=~8 and lower (on our $k$~=~3 parameters; 
i.e., for water pressure or water content, surface gravity, and stellar flux), 
reaching a total of $N$~=~$\dbinom{3+8}{8}$~=~165 polynomial terms for $n$~=~8.

As a third step, we used the recursive feature elimination (RFE) iterative method \citep{Scikit:2011} 
to derive the optimal polynomial fit of our data, using the RFE python tool\footnote{\url{https://scikit-learn.org/stable/modules/generated/sklearn.feature_selection.RFE.html}}.
The RFE method was implemented following the recursive steps described below, for each $n$ (from $n$~=~8 to 1):
\begin{enumerate}
 \item We performed a linear fit of our modeled data (surface temperature and effective temperature) 
with the polynomial of $N$~terms (initially, and for reference, $N$~=~$\frac{(k+n)!}{n!~k!}$).
 \item We calculated the RMS of the fit. 
 \item We evaluated the absolute contribution of all $N$ polynomial terms to the fit.
 \item We removed the polynomial term with the smallest absolute contribution.
 \item We restarted the procedure iteratively with $N$-1~terms, and until $N$~=~1.
\end{enumerate}

\begin{figure*}
    \centering
\includegraphics[width=\linewidth]{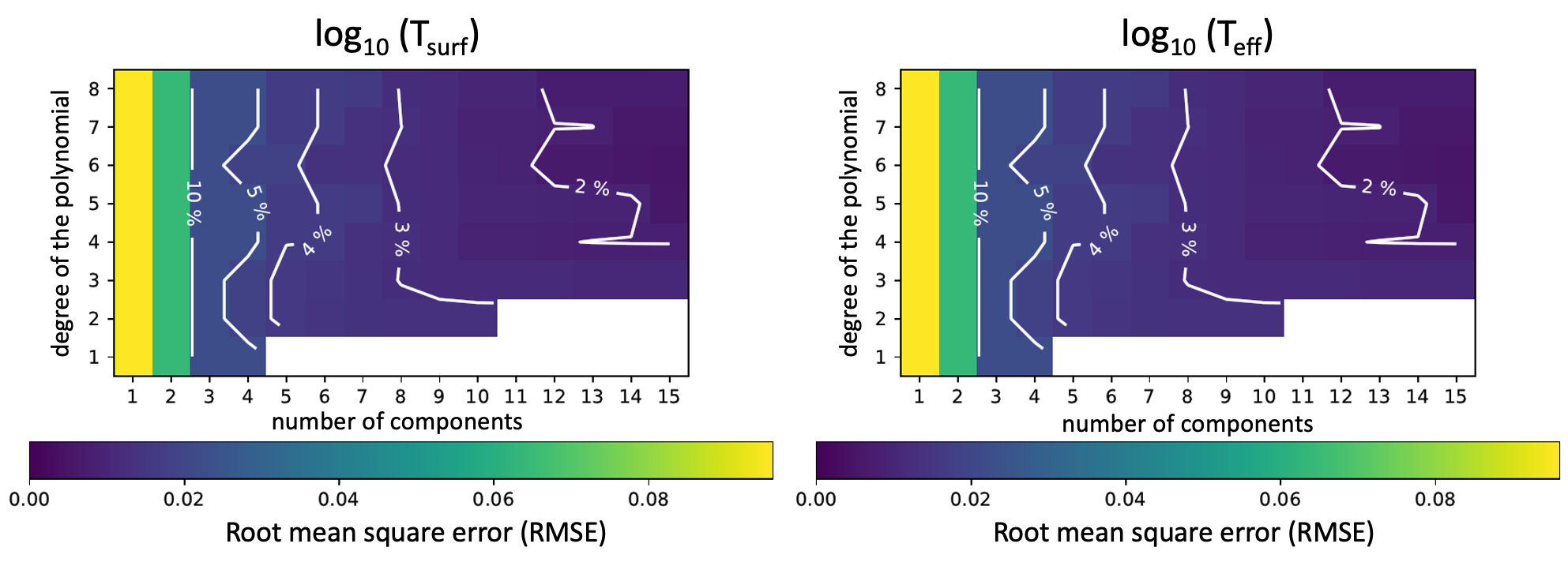}
\caption{Maps of the root mean square error (RMSE) for the fits 
on the surface temperature T$_{\text{surf}}$ (left panel) 
and the effective temperature T$_{\text{eff}}$ (right panel), as a function of the 
number of polynomial components and the initial degree of the polynomial.}
\label{figure_fit1}
\end{figure*}

As the final step, we compared the RMS of the fit for each polynomial 
in order to derive the best compromise between the value of the RMS and the number $N$ of polynomial terms. 
Based on Fig.~\ref{figure_fit1}, we decided that the fit that gives the best compromise is found for $N$~=~10 
for both the surface temperature and the effective temperature.
Fig.~\ref{figure_fit1} shows the distribution of the residuals of the fit for the surface 
temperature and the effective temperature, 
thus making it possible to evaluate the goodness of the fit (mean error $\sim$~2.5$\%$; max error $\sim$~10$\%$).

\begin{figure*}
    \centering
\includegraphics[width=\linewidth]{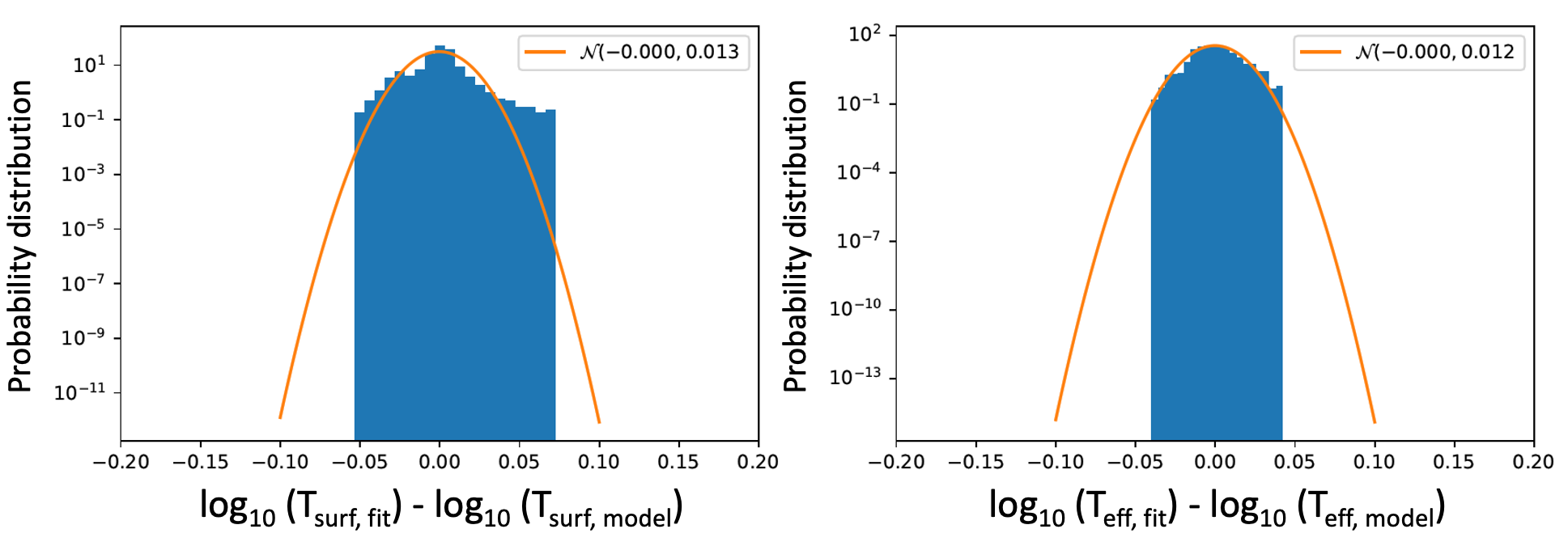}
\caption{Probability distributions (blue histogram) of residuals for the fits 
on the surface temperature T$_{\text{surf}}$ (left panel) 
and the effective temperature T$_{\text{eff}}$ (right panel). A total of 16488 1-D numerical 
atmospheric simulations were used. The orange curves indicate 
the normal distribution laws that best fit the distributions.}
\label{figure_fit2}
\end{figure*}

\section{Procedure to derive the empirical formula of the thickness of a steam H$_2$O atmosphere}
\label{appendix_fit_thickness}

To construct Equation~\ref{empirical_equation_thickness}, 
we first assumed the hydrostatic equilibrium:
\begin{equation}
dP+\rho \text{g} dr~=~0,
\end{equation} 
with P the atmospheric pressure, r the radial coordinate, g the gravity, and r the radial coordinate. 
g can be written as
\begin{equation}
\text{g}(r)~=~\text{g}_{\text{core}}~\times~(\frac{\text{R}_{\text{core}}^2}{r^2}).
\end{equation}
We then assumed the atmosphere follows the perfect gas law:
\begin{equation}
\rho~=~\frac{P~\text{M}_{\text{H}_2\text{O}}}{\text{R}~\text{T}_\text{eff}}
,\end{equation} 
with R the gas constant and M$_{\text{H}_2\text{O}}$ the molecular weight of H$_2$O (here the dominant gas).
T$_{\text{eff}}$ is the effective atmospheric temperature and assumed to be constant, for simplicity.

Combining the three previous equations, we derived:
\begin{equation}
d(\log{P})~=~(\frac{\text{M}_{\text{H}_2\text{O}}~\text{g}_{\text{core}}}{\text{R}~\text{T}_\text{eff}})~\text{R}_{\text{core}}^2~d(\frac{1}{r}).
\end{equation} 
We then integrated this equation (assuming P~=~P$_{\text{surf}}$ at r~=~R$_{\text{core}})$:
\begin{equation}
\log{(\frac{P}{P_\text{surf}})}~=~(\frac{\text{M}_{\text{H}_2\text{O}}~\text{g}_{\text{core}}}{\text{R}~\text{T}_\text{eff}})~\times~\text{R}_{\text{core}}^2~\times~(\frac{1}{r}-\frac{1}{\text{R}_{\text{core}}}).
\end{equation} 

At the transit radius, R~=~R$_{\text{p}}$ and P~=~P$_{\text{transit}}$, which gives
\begin{equation}
\log{(\frac{P_\text{transit}}{P_\text{surf}})}~=~(\frac{\text{M}_{\text{H}_2\text{O}}~\text{g}_{\text{core}}}{\text{R}~\text{T}_\text{eff}})~\times~\text{R}_{\text{core}}^2~\times~(\frac{1}{\text{R}_{\text{p}}}-\frac{1}{\text{R}_{\text{core}}}),
\end{equation} 
which can be rewritten as
\begin{equation}
\text{R}_{\text{p}}~=~\Big((\frac{\text{R}~\text{T}_\text{eff}}{\text{M}_{\text{H}_2\text{O}}~\text{g}_{\text{core}}})~\times~(\frac{1}{\text{R}_{\text{core}}^2})~\times~\log{(\frac{P_\text{transit}}{P_\text{surf}})}+\frac{1}{\text{R}_{\text{core}}}\Big)^{-1.}
\end{equation}
With R$_{\text{p}}$~=~R$_{\text{core}}$+z$_{\text{atmosphere}}$, we have
\begin{equation}
z_{\text{atmosphere}}~=~\text{R}_{\text{core}}~\Big(\frac{\text{R}_{\text{core}}}{\log{(\frac{P_\text{surf}}{P_\text{transit}})}~(\frac{\text{R}~\text{T}_\text{eff}}{\text{M}_{\text{H}_2\text{O}}~\text{g}_{\text{core}}})}-1\Big)^{-1.}
\end{equation}

We then assumed
\begin{equation}
 P_\text{surf}~=~\frac{\text{M}_\text{atmosphere}~\text{g}_{\text{core}}}{4\pi~\text{R}_{\text{core}}^2}
,\end{equation} 
with M$_\text{atmosphere}$ the mass of the steam H$_2$O-dominated atmosphere (in kg). 
This relationship does not hold for inflated atmospheres, but for simplicity, we assumed it is valid anyway. 
Moreover, we have
\begin{equation}
\text{M}_\text{atmosphere}~=~\frac{\text{M}_\text{core}~\text{x}_{\text{H}_2\text{O}}}{1-\text{x}_{\text{H}_2\text{O}}}~=~\frac{\text{g}_\text{core}~\text{R}_\text{core}^2}{G}~\times~(\frac{\text{x}_{\text{H}_2\text{O}}}{1-\text{x}_{\text{H}_2\text{O}}}).
\end{equation} 

Combining the three previous equations leads to Equation~\ref{empirical_equation_thickness}. 

\section{Quick guide on how to build mass-radius relationships for water-rich rocky planets more irradiated than the runaway greenhouse limit.}
\label{appendix_how_to}

In this Appendix, we provide a procedure that can be followed 
to build mass-radius relationships for water-rich rocky planets more irradiated than 
the runaway greenhouse limit:
\begin{enumerate}
 \item Choose a core composition. 
 \item Retrieve (or calculate) the mass-radius relationship corresponding to this core composition. For instance, 
\citet{Zeng:2016}\footnote{User-friendly data is provided on the personal website 
of Li Zeng (\url{https://www.cfa.harvard.edu/~lzeng/planetmodels.html})}
provides ascii tables of mass-radius relationships for a wide range of interior composition.
 \item Choose the water mass fraction (x$_{\text{H}_2\text{O}}$) of your planets, 
as well as the irradiation (S$_{\text{eff}}$) they receive. We note that the irradiation must be 
larger than the runaway greenhouse irradiation limit, which depends on the type of host star \citep{Kopparapu:2013} 
and on the mass and radius of the planetary core \citep{Kopparapu:2014}. Moreover, 
the water mass fraction must be "reasonable" (see discussions in Appendix~\ref{appendix_trick_r-m}).
\item For each datapoint of the selected core mass-radius relationship (i.e., for each set of core mass and radius), 
calculate the corresponding surface gravity ($g_{\text{core}}$).
\item For each datapoint of the selected core mass-radius relationship, 
compute the thickness z$_{\text{atmosphere}}$ of the H$_2$O atmospheric layer using Equation~\ref{empirical_equation_thickness}. 
Equation~\ref{empirical_equation_thickness} makes use of Equation~\ref{empirical_equation_teff} and the empirical 
coefficients provided in Table~\ref{tab:table_coeff_teff}.
\item For each datapoint of the selected core mass-radius relationship (M$_{\text{core}}$,R$_{\text{core}}$), 
compute the new mass-radius relationship (M$_{\text{planet}}$,R$_{\text{planet}}$) 
by assuming that R$_{\text{planet}}$~=~R$_{\text{core}}$+z$_{\text{atmosphere}}$ and 
M$_{\text{planet}}$~=~M$_{\text{core}}$~/~(1-x$_{\text{H}_2\text{O}}$).
\end{enumerate}

\end{document}